\documentclass[conference]{IEEEtran}
\usepackage{epsfig}
\usepackage{amssymb}
\usepackage{latexsym,graphicx}
\usepackage{longtable}
\usepackage{algpseudocode} 
\usepackage{algorithm}
\usepackage{array}
\usepackage{mathrsfs}
\usepackage{color}
\usepackage{amsmath,graphicx}
\usepackage{epsfig}
\usepackage{amssymb}
\usepackage{setspace}
\usepackage{cite}
\usepackage{hyperref}
\usepackage{amssymb}
\usepackage{caption}
\usepackage{subcaption}
\def\BibTeX{{\rm B\kern-.05em{\sc i\kern-.025em b}\kern-.08em
    T\kern-.1667em\lower.7ex\hbox{E}\kern-.125emX}}
 \makeatletter
 \let\NAT@parse\undefined
 \makeatother

\begin{document}

\title{\huge Transmission Capacity of Full-Duplex MIMO Ad-Hoc Network with Limited Self-Interference Cancellation}
%
%
\author{$\text{Jiancao Hou}$ and $\text{Mohammad Shikh-Bahaei}$
\\{\small Centre for Telecommunications Research, King's College London, London, UK}
\\{\small e-mail:\{jiancao.hou, m.sbahaei\}@kcl.ac.uk}


}
\markboth{} {Shell \MakeLowercase{\textit{et al.}}: Bare Demo of
IEEEtran.cls for Journals}\maketitle

\begin{abstract}
In this paper, we propose a joint transceiver beamforming design to simultaneously mitigate self-interference (SI) and partial inter-node interference for full-duplex multiple-input and multiple-output ad-hoc network, and then derive the transmission capacity upper bound (TC-UB) for the corresponding network. Condition on a specified transceiver antenna's configuration, we allow the SI effect to be cancelled at transmitter side, and offer an additional degree-of-freedom at receiver side for more inter-node interference cancellation. In addition, due to the proposed beamforming design and imperfect SI channel estimation, the conventional method to obtain the TC-UB is not applicable. This motivates us to exploit the dominating interferer region plus Newton-Raphson method to iteratively formulate the TC-UB. The results show that the derived TC-UB is quite close to the actual one especially when the number of receive-antenna is small. Moreover, our proposed beamforming design outperforms the existing beamforming strategies, and FD mode works better than HD mode in low signal-to-noise ratio region. 
\end{abstract}


\IEEEpeerreviewmaketitle
 

\section{Introduction}
Due to rapid expansion of applications and services, next generation wireless networks are expected to rely on low latency and high spectral efficiency \cite{Orikumhi2017,Towhidlou2018}. In-band full-duplex (IBFD) as one of promising techniques has recently re-emerged to achieve such requirements \cite{Sabharwal2014,Kim2015,Chen2017,Naslcheraghi2017}. Unlike the conventional half-duplex (HD) radio transceiver design, IBFD is able to transmit and receive simultaneously over the same spectrum. In theory, it can halve latency and double spectral efficiency of point-to-point communications if the self-interference (SI) effect can be cancelled perfectly. On the other hand, many researchers have contributed to develop advanced techniques to handle the SI effect, such as propagation-domain antenna separation \cite{Choi2010,Everett2014}, analog-domain and/or digital-domain SI cancellation \cite{Duarte2010,Jain2011,Duarte2012}. Although these techniques can potentially yield impressive SI cancellation, their performances degrade considerably if the SI channel state information (CSI) is not modelled/estimated accurately. 

Consider a FD ad-hoc wireless network, where multiple transceiver pairs communicate simultaneously without central control unit. The SI and the inter-node interference as the significant barriers limit the rate of successful transmissions. Employing multiple-input and multiple-output (MIMO) at each node is a way to manage these interferences and improve the successful transmissions with high data rate \cite{Mohammadi2015,Huberman2015}. On the other hand, transmission capacity (TC), which characterizes the maximum node density of successful transmissions of a network, has been investigated intensively for some specific HD MIMO strategies in \cite{Hunter2008,Vaze2012,Huang2012}. Since there is no SI presented, the performance analysis in terms of TC can be easily conducted by exploiting the statistical property of the accumulated inter-node interference via 1-D Poisson point process (PPP). Considering FD with multiple antennas at each node, the SI has attracted many research efforts to cancel it in spatial domain \cite{Riihonen2011,Senaratne2011}. In addition, the performance analysis on the network have been investigated in \cite{Ju2012,Psomas2017,Mungara2017,Atzeni2017}, where the authors provided the throughput and/or achievable sum-rate analysis in the presence of residual SI and with some inter-node interference cancellation strategies. To the best of our knowledge, the theoretical analysis of compact TC upper bound (TC-UB) for FD MIMO ad-hoc network in the presence of SI channel estimation error has not been well studied, where the analysed TC framework can give a simple expression for quantification of achievable rate.

Motivated by the above discussion, in this paper, we first propose a joint transceiver beamforming design to mitigate SI and partial inter-node interference effects for the FD MIMO ad-hoc network. In this case, by exploiting a specific transceiver antenna's configuration, the estimated SI can be cancelled at transmitter side and leave an additional degree-of-freedom (DoF) at receiver side for more inter-node interference cancellation. Then, we derive the compact TC-UB of considered model in the presence of SI channel estimation error. Unlike the TC-UB derivation in HD case \cite{Hunter2008,Vaze2012,Huang2012}, the TC-UB derivation in FD case cannot resort to the standard 1-D PPP to get the expected distance of dominating interferer from the typical receiver, due to SI channel estimation error. Thus, we exploit the dominating interferer region with generalized 2-D PPP and then utilize Newton-Raphson method to iteratively formulate the TC-UB. The results show that the derived TC-UB is quite close to the simulated curve especially with large transmit-antenna to receive-antenna ratio (TRR). Moreover, our proposed transceiver beamforming design outperforms the existing beamforming strategies, e.g., singular value decomposition (SVD) and partial zero-forcing (ZF) methods. Finally, we compare our derived TC-UB in FD mode with the ones in HD mode, and find the break-even point, where FD and HD provide the same TC performance.




\section{System Model}
Consider an interference-limited ad-hoc network, where each node has FD capability and is equipped with $N$ antennas. 
$N_t$ is the number of transmit-antenna per node, and $N_r(=N-N_t)$ is the number of receive-antenna per node. The nodes in the network are divided into two different sets, e.g., $\Phi_{A}=\{a_{k,k=0,1,\ldots,M-1}\}$ and $\Phi_{B}=\{b_{k,k=0,1,\ldots,M-1}\}$. The nodes locations of set $\Phi_{A}$ follows a stationary PPP with intensity $\lambda$, and the nodes locations of set $\Phi_{B}$ can be written as $b_{k}=a_{k}+L_{k}(\cos\varphi_{k},\sin\varphi_{k}),\forall k$, where $L_{k}$ is the distance between $a_{k}$ and $b_{k}$, and $\varphi_{k}$ is the angle following independent and uniformly distributed on $[0,2\pi]$. To simplify the analysis, the distance of each transceiver node pair is fixed to $L$, and the transmission power of each node is set to $P$. We also assume each node can perfectly estimate its associated links' CSIs except the SI channel.

We select a node pair, e.g., $\{a_{0},b_{0}\}$, as the typical link pair. According to Slivnyak's theorem \cite{Haenggi2012}, the properties of node $a_0\in\Phi_{A}$ (or node $b_0\in\Phi_{B}$) represent the properties of other nodes in the same set. In addition, due to FD capability, node $a_0$ receives interferences not only from $\Phi_{A}/a_0$ but also from $\Phi_{B}/b_0$. Accordingly, the received signal at node $i\neq j\in\{a_0,b_0\}$ with size of $N_r\times1$ can be expressed as
\begin{IEEEeqnarray}{ll}\label{eq01}
\mathbf{y}_{i}&=\sqrt{P}L^{-\alpha/2}\mathbf{H}_{i,j}\mathbf{x}_{j}+\sqrt{P}\sum^{M-1}_{k=1}r^{-\alpha/2}_{b_k}\mathbf{H}_{i,b_k}\mathbf{x}_{b_k}\nonumber\\
&~~+\sqrt{P}\sum^{M-1}_{k=1}r^{-\alpha/2}_{a_k}\mathbf{H}_{i,a_k}\mathbf{x}_{a_k}+\sqrt{P}\mathbf{H}_{i}\mathbf{x}_{i}+\mathbf{v}_{i},
\end{IEEEeqnarray}
where $\mathbf{x}_{i}$ and $\mathbf{x}_{j}$ are with size of $N_t\times1$ are the transmitted signals from the typical node $i$ and $j$, respectively; $\mathbf{x}_{b_k}$ and $\mathbf{x}_{a_k}$ with size of $N_t\times1$ are the transmitted signals from node $b_k$ and node $a_{k}$, respectively; $\mathbf{H}_{i,b_k}$ and $\mathbf{H}_{i,a_k}$ with size of $N_r\times N_t$ denote channel fading effects from node $b_k$ to node $i$ and from node $a_k$ to node $i$, respectively, and they are independent and identically distributed (i.i.d.) complex Gaussian random variables with zero mean and unit variance. $r_{b_k}$ and $r_{a_k}$ represent the distances from node $b_k$ to node $i$ and from node $a_k$ to node $i$, respectively. $\alpha>2$ is the path-loss exponent. $\mathbf{v}_{i}$ with size of $N_r\times1$ denotes the additive white Gaussian noise (AWGN) at node $i$ with zero mean and unit variance. Moreover, due to imperfect SI channel estimation, we have
\begin{equation}\label{eq02}
\mathbf{H}_{i} = \hat{\mathbf{H}}_{i} + \mathbf{\Delta}_{i},
\end{equation}
where $\mathbf{H}_{i}$ is the actual self-interference channel, $\hat{\mathbf{H}}_{i}$ is its estimated version, and $\mathbf{\Delta}_{i}$ is the estimation error following complex Gaussian distribution with zero mean and $\sigma^{2}_{i,\mathrm{SI}}$ variance. Here, we assume $\hat{\mathbf{H}}_{i}$ and the statistical information of $\mathbf{\Delta}_{i}$ are known by node $i$.  

With the local CSIs at each node, we can formulate the pre-processing/post-processing beamforming vectors to mitigate the SI and the inter-node interference. Specifically, let's denote $\mathbf{w}_{b_k}\in\mathcal{C}^{N_{t}\times1}$ and $\mathbf{w}_{a_k}\in\mathcal{C}^{N_{t}\times1}$ as the pre-processing beamforming vectors for node $b_k$ and node $a_k$, respectively. Then, we have the transmitted signals
\begin{equation}\label{eq03}
\mathbf{x}_{b_k}=\mathbf{w}_{b_k}s_{b_k},~k=0,1,\ldots,M-1,
\end{equation}
\begin{equation}\label{eq04}
\mathbf{x}_{a_k}=\mathbf{w}_{a_k}s_{a_k},~k=0,1,\ldots,M-1,
\end{equation}
where $s_{b_k}$ and $s_{a_k}$ represent the data symbols at node $b_k$ and node $a_k$, respectively, and the power of $\mathbf{x}_{b_k}$ and $\mathbf{x}_{a_k}$ are normalized to one. Subsequently, the received signal at the typical node $i\neq j\in\{a_0,b_0\}$, after applying the post-processing beamforming vector $\mathbf{z}_{i}$, can be expressed as
\begin{IEEEeqnarray}{ll}\label{eq05}
\mathbf{z}^{\mathrm{H}}_{i}\mathbf{y}_{i}&=\sqrt{P}L^{-\alpha/2}h_{i,j}+\sqrt{P}\sum^{M-1}_{k=1}r^{-\alpha/2}_{b_k}h_{i,b_k}\nonumber\\
&~~+\sqrt{P}\sum^{M-1}_{k=1}r^{-\alpha/2}_{a_k}h_{i,a_k}+\sqrt{P}h_i+v_i,
\end{IEEEeqnarray}
where $h_{i,j}\triangleq\mathbf{z}^{\mathrm{H}}_{i}\mathbf{H}_{i,j}\mathbf{x}_{j}$, $h_{i,b_k}\triangleq\mathbf{z}^{\mathrm{H}}_{i}\mathbf{H}_{i,b_k}\mathbf{x}_{b_k}$, $h_{i,a_k}\triangleq\mathbf{z}^{\mathrm{H}}_{i}\mathbf{H}_{i,a_k}\mathbf{x}_{a_k}$, $h_i\triangleq\mathbf{z}^{\mathrm{H}}_{i}\mathbf{H}_{i}\mathbf{x}_{i}$, $v_i\triangleq\mathbf{z}^{\mathrm{H}}_{i}\mathbf{v}_{i}$, and $(\cdot)^{H}$ denotes conjugate transpose. Then, the performance metric signal-to-interference plus noise ratio (SINR) at node $i$ is given by
\begin{IEEEeqnarray}{ll}\label{eq06}
\mathrm{SINR}_{i}=\frac{L^{-\alpha}|h_{i,j}|^2}{\sum^{M-1}_{k=1}(r^{-\alpha}_{b_k}|h_{i,b_k}|^2+r^{-\alpha}_{a_k}|h_{i,a_k}|^2)+|h_i|^2+\frac{|v_{i}|^2}{P}},\nonumber\\
\end{IEEEeqnarray}
where $|\cdot|$ denotes absolute value.



\section{Transceiver Beamforming Design}
In this section, we introduce an efficient transceiver beamforming design to balance desired signal's power and co-channel interference cancellation. With fixed number of transceiver antennas per node, we formulate the pre-processing/post-processing beamforming vectors at each node to solve the following optimization problem
\begin{IEEEeqnarray}{ll}\label{eq07}
\underset{\mathbf{z}_{i},\mathbf{w}_{i},\mathbf{w}_{b_k},\mathbf{w}_{a_k}}{\mathrm{maximize}}&~~|\mathbf{z}^{H}_{i}\mathbf{H}_{i,j}\mathbf{w}_{j}|^{2}\\
~~\mathrm{subject~to}&~~|\mathbf{z}^{H}_{i}\mathbf{H}_{i,b_k}\mathbf{w}_{b_k}|^2=0,~k=1,\ldots,{l}/{2},\nonumber\\
&~~|\mathbf{z}^{H}_{i}\mathbf{H}_{i,a_k}\mathbf{w}_{a_k}|^2=0,~k=1,\ldots,l/2,\nonumber\\
&~~|\mathbf{z}^{H}_{i}\hat{\mathbf{H}}_{i}\mathbf{w}_{i}|^2=0,~i\neq j\in\{a_0,b_0\},\nonumber
\end{IEEEeqnarray}
where the receiver $i$ aims to cancel the inter-node interference from $l/2$ nearest transceiver node pairs. To obtain the optimal beamforming vector solution, all the nodes need to chase each other's transceiver beams and jointly make the final decision, which is quite challenge and at cost of heavy signalling overhead. This motivates us to design the beamforming vectors in a distributed manner. 

By assuming all channel matrices come with full rank, the design criteria follows two conditions: 1) $N_t>N_r$; 2) $N_t\leq N_r$. For the condition that $N_t>N_r$, the SI effect, i.e., $\mathbf{z}^{H}_{i}\hat{\mathbf{H}}_{i}\mathbf{w}_{i}, _{i\in\{a_0,b_0\}}$, can be nullified by $\mathbf{w}_{i}$ itself irrespective of $\mathbf{z}_{i}$. Then, the node $i$ can spend all its $N_r-1$ spatial DoFs at receiver side to cancel $\lfloor\frac{N_r-1}{2}\rfloor$ nearest interference pairs. 
In detail, let's first decompose $\hat{\mathbf{H}}_{i}$ into spatial modes by SVD as $\hat{\mathbf{H}}_{i}=\mathbf{U}_{i}\mathbf{\Sigma}_{i}\mathbf{V}^{H}_{i}$. Then, the ($\mathrm{rank}(\hat{\mathbf{H}}_{i})+1)$ to $N_t$ column(s) of $\mathbf{V}_{i}$ span null space of $\hat{\mathbf{H}}_{i}$ and construct a matrix $\tilde{\mathbf{V}}_{i}$ (or a vector $\tilde{\mathbf{v}}_{i}$) with size of $N_t\times(\mathrm{rank}(\hat{\mathbf{H}}_{i})+1:N_t)$. Then, to improve transmission quality of the desired link, i.e., $\mathbf{H}_{j,i},_{j\neq i\{a_0,b_0\}}$, we need to define $\tilde{\mathbf{H}}_{j,i}\triangleq\mathbf{H}_{j,i}\tilde{\mathbf{V}}_{i}$, and formulate the SVD of $\tilde{\mathbf{H}}_{j,i}$ as $\tilde{\mathbf{U}}_{j,i}\tilde{\mathbf{\Sigma}}_{j,i}\tilde{\mathbf{V}}^{H}_{j,i}$. Then, the pre-processing beamforming vector at node $i$ can be formulated as
\begin{equation}\label{eq08}
\mathbf{w}_{i}=\tilde{\mathbf{V}}_{i}\tilde{\mathbf{v}}^{(1)}_{j,i},~_{i\neq j\in\{a_0,b_0\}},
\end{equation} 
where $\tilde{\mathbf{v}}^{(1)}_{j,i}$ denotes the first column of $\tilde{\mathbf{V}}_{j,i}$. Similar way to formulate pre-processing beamforming vectors for all nodes in $\Phi_{A}/a_0$ and $\Phi_{B}/b_0$. Correspondingly, the post-processing beamforming vector $\mathbf{z}_{j},_{j\neq i\in\{a_0,b_0\}}$ at node $j$ can be formulated as 
\begin{equation}\label{eq09}
\mathbf{z}_{j}=(\mathbf{s}^{H}_{j}\tilde{\mathbf{u}}^{(1)}_{j,i})^{-1}\mathbf{s}^{H}_{j},~_{j\neq i\in\{a_0,b_0\}},
\end{equation}
where $\tilde{\mathbf{u}}^{(1)}_{j,i}$ denotes the first column of $\tilde{\mathbf{U}}_{j,i}$, and $\mathbf{s}_{j}$ with size of $N_r\times1$ is used to span null space of $\mathbf{H}_{j,b_k}\mathbf{w}_{b_k}$ and $\mathbf{H}_{j,a_k}\mathbf{w}_{a_k}$, $k=1,\ldots,l/2$. 

For the condition that $N_t\leq N_r$, $\hat{\mathbf{H}}_{i}$ has full column rank. In this case, we have to cancel the SI plus $\lfloor\frac{N_r-2}{2}\rfloor$ nearest interference pairs at receiver side. Thus, to improve transmission quality of the desired link, we formulate the pre-processing beamforming vector $\mathbf{w}_{i}$ at node $i$ as
\begin{equation}\label{eq10}
\mathbf{w}_{i}=\mathbf{v}^{(1)}_{j,i},~_{i\neq j\in\{a_0,b_0\}},
\end{equation} 
where $\mathbf{v}^{(1)}_{j,i}$ is the first right singular vectors of channel matrix $\mathbf{H}_{j,i}$. Similar way to formulate pre-processing beamforming vectors for all nodes in $\Phi_{A}/a_0$ and $\Phi_{B}/b_0$. Correspondingly, the post-processing beamforming vector $\mathbf{z}_{j},_{j\neq i\in\{a_0,b_0\}}$ at node $j$ can be formulated as 
\begin{equation}\label{eq11}
\mathbf{z}_{j}=(\mathbf{s}^{H}_{j}\mathbf{u}^{(1)}_{j,i})^{-1}\mathbf{s}^{H}_{j},~_{j\neq i\in\{a_0,b_0\}},
\end{equation}
where $\mathbf{u}^{(1)}_{j,i}$ denotes the first left singular vectors of channel matrix $\mathbf{H}_{j,i}$, and $\mathbf{s}_{j}$ with size of $N_r\times1$ in this case is used to span null space of $\hat{\mathbf{H}}_{j}\mathbf{w}_{j}$, $\mathbf{H}_{j,b_k}\mathbf{w}_{b_k}$ and $\mathbf{H}_{j,a_k}\mathbf{w}_{a_k}$, $k=1,\ldots,l/2$.\footnote{In this paper, we assume the SI always has priority to be cancelled first. More detailed analysis of why to cancel SI at first can be found in \cite{Atzeni2017}.} 

With above discussed transceiver beamforming vectors design, the received signal at typical node $i\neq j\in\{a_0,b_0\}$ can be expressed as
\begin{IEEEeqnarray}{ll}
\mathbf{z}^{\mathrm{H}}_{i}\mathbf{y}_{i}&=\sqrt{P}L^{-\alpha/2}\sqrt{\gamma_{i,j}}s_{j}+\sqrt{P}\sum^{M-1}_{k=\frac{l}{2}+1}r^{-\alpha/2}_{b_k}h_{i,b_k}\nonumber\\
&~~+\sqrt{P}\sum^{M-1}_{k=\frac{l}{2}+1}r^{-\alpha/2}_{a_k}h_{i,a_k}+\mathbf{z}^{\mathrm{H}}_{i}\sqrt{P}\mathbf{\Delta}_{i}\mathbf{x}_{i}+v_{i},
\end{IEEEeqnarray}
where $\gamma_{i,j}$ is the largest eigenvalue of $\tilde{\mathbf{H}}_{i,j}\tilde{\mathbf{H}}^{H}_{i,j}$ for the condition that $N_t>N_r$, and the largest eigenvalue of $\mathbf{H}_{i,j}\mathbf{H}^{H}_{i,j}$ for the condition that $N_t\leq N_r$. Then, the corresponding SINR can be formulated as
\begin{IEEEeqnarray}{ll}\label{eq13}
\mathrm{SIR}_{i}=\frac{L^{-\alpha}\gamma_{i,j}}{\sum^{M-1}_{k=\frac{l}{2}+1}r^{-\alpha}_{k}(|h_{i,b_k}|^2+|h_{i,a_k}|^2)+|h^{'}_i|^2+\frac{|v_{i}|^2}{P}},\nonumber\\
\end{IEEEeqnarray}
where $h^{'}_{i}\triangleq\mathbf{z}^{\mathrm{H}}_{i}\mathbf{\Delta}_{i}\mathbf{x}_{i}$ is the residual SI effect due to channel estimation error. Here, we assume $r_k\triangleq r_{b_k}=r_{a_k},\forall k$, where the interference power from each interfering pair is measured from the same distance but with independent channel fading. Such model is widely used to analyse the  scaling properties of average sum-rate as in \cite{Ju2012}, or analyse upper bound of network success probability as in \cite{Tong2015}. 

\section{Transmission Capacity Analysis}
There is a probability that high spatial throughput is obtained accompanied with unacceptably high outage, which results in a large number of wasted transmissions. This motivates TC to be proposed to exploit the number of successful transmissions in a unit area with a permissible outage constraint \cite{Weber2010}. We assume that the OP, i.e., $q(\lambda)$, is a function of user density $\lambda$. Then, the TC formula is given by \cite{Weber2010}
\begin{equation}\label{eq14}
c(\epsilon)=q^{-1}(\epsilon)(1-\epsilon)R,~\epsilon\in(0,1),
\end{equation}
where $\epsilon$ is a target OP, and $R=\log_{2}(1+\beta)$ is a target transmission rate with SINR threshold $\beta$. In general, it is quite difficult to compute the exact closed-form for the OP/TC due to the unavailability of closed-form expressions for the largest eigenvalue of desired channel link and the distribution of sum of the remaining interferers' distances from the typical receiver. This motivates to exploit the compact expression of the lower/upper bounds for OP/TC.

\subsection{OP Lower Bound \& TC Upper Bound}
The OP-LB can be obtained by introducing the dominating interferer pair(s), i.e., the $D$ nearest interference pair(s) from the typical receiver after the proposed interference cancellation, which is
\begin{eqnarray}\label{eq15}
q(\epsilon)&=&\mathrm{Pr}\left[\mathrm{SINR}_{i}<\beta\right]\nonumber\\
&>&\mathrm{Pr}\left[\frac{1}{\mathrm{SINR}^{(d)}_{i}}>\frac{1}{\beta}\right],
\end{eqnarray}
where
\begin{equation}
\mathrm{SINR}^{(d)}_{i}\triangleq\frac{L^{-\alpha}\gamma_{i,j}}{\sum^{\frac{l}{2}+D}_{k=\frac{l}{2}+1}r^{-\alpha}_{k}\left(|h_{i,b_{k}}|^2+|h_{i,a_{k}}|^2\right)+|h^{'}_i|^2+\frac{|v_i|^2}{P}}\nonumber
\end{equation}
is the SINR consisted of the $D$ dominating interferer pair(s) after the interference cancellation. Here, the dominating interferer pair is defined as its interference contribution alone plus the SI channel estimation error and noise are sufficient to cause outage at the typical receiver. Then, the geometrical location region of the dominating interferer can be obtained as
\begin{eqnarray}\label{eq16}
R_d&\triangleq &\left\{r_{k}:\mathrm{SINR}^{(1)}_{i}=\beta\right\}\nonumber\\
&=&\left[\frac{\beta\left(|h_{i,b_{k}}|^2+|h_{i,a_{k}}|^2\right)}{L^{-\alpha}\gamma_{i,j}-\beta (|h^{'}_i|^2+\frac{|v_i|^2}{P})}\right]^{\frac{1}{\alpha}},
\end{eqnarray}
where $R_d$ is the radius from the typical receiver of the dominating interferer region (e.g. $\mathcal{I}$). In this case, any node pair after the interference cancellation process located in this region $\mathcal{I}$ is deemed as the dominating interferer pair. Thus, the OP-LB in \eqref{eq15} can be obtained by exploiting its equivalent event that at least one interference pair is located within the dominating interferer region $\mathcal{I}$, which is  
\begin{eqnarray}\label{eq17}
q^{l}(\lambda)&=&\mathrm{Pr}\left[r_{\frac{l}{2}+1}\leq R_d\right]\nonumber\\
&\overset{(a)}{=}&\mathbb{E}_{S}\left\{\mathbb{E}_{SI}\left\{\mathbb{E}_{\psi}\left\{\frac{\gamma(\frac{l}{2}+1,\lambda\pi R^{2}_d)}{\Gamma(\frac{l}{2}+1)}\right\}\right\}\right\},
\end{eqnarray}
where $\gamma(\cdot,\cdot)$ is the lower incomplete gamma function, $\Gamma(\cdot)$ is the gamma function, $\mathbb{E}_{S}\{\cdot\}$ is the expectation with respect to (w.r.t.) $\gamma_{i,j}$, $\mathbb{E}_{SI}\{\cdot\}$ is the expectation w.r.t. $|h^{'}_i|^2$, and $\mathbb{E}_{\psi}\{\cdot\}$ is the expectation w.r.t. $\psi_{i,k}\triangleq|h_{i,b_{k}}|^2+|h_{i,a_{k}}|^2$. In addition, Step (a) in \eqref{eq17} is obtained by formulating the cumulative distribution function of Euclidean distance between the $(\frac{l}{2}+1)^{\mathrm{th}}$ interference pair and the typical receiver as shown in \cite{Haenggi2005}.

To obtain TC-UB, we first need to obtain $\lambda$ by inverting the OP-LB $q^{l}(\lambda)$ in \eqref{eq17}. However, such operation is infeasible due to the expectations in front of the low incomplete gamma function. On the other hand, with the condition that $a\leq b+1$, $\gamma(a,b)$ is a convex function. In this case, following Jensen's inequality, the OP-LB can be approximated as
\begin{equation}\label{eq18}
q^{l}(\lambda)\approx \frac{\gamma(\frac{l}{2}+1,\lambda\pi\Omega)}{\Gamma(\frac{l}{2}+1)},
\end{equation}
where
\begin{equation}\label{eq19}
\Omega\triangleq\beta^{\frac{2}{\alpha}}\mathbb{E}\{\psi_{i,k}^{\frac{2}{\alpha}}\}(\frac{\mathbb{E}\{\gamma_{i,j}\}}{L^{\alpha}}-\beta \mathbb{E}\{|h^{'}_i|^2\})^{-\frac{2}{\alpha}}.
\end{equation}
Here, we should have $\frac{l}{2}\leq\lambda\pi\Omega$. Then, following Taylor expansion of the lower incomplete Gamma function as in  \cite{Abramowitz1970}, \eqref{eq18} can be converted to
\begin{eqnarray}\label{eq20}
q^{l}(\lambda)=\frac{(\pi\Omega)^{\frac{l}{2}+1}}{\Gamma(\frac{l}{2}+1)}\sum^{\infty}_{m=0}\frac{(-\pi\Omega)^{m}}{m!}\frac{\lambda^{m+\frac{l}{2}+1}}{m+\frac{l}{2}+1}.
\end{eqnarray}
Then, to find $\lambda$ such that $q^{l}(\lambda)=\epsilon$, we can resort to Newton-Raphson method \cite{Suli2003}. Specifically, let's start from an initial guess $\lambda_{0}$ for a root of the function $q^{l}(\lambda)=\epsilon$, which is 
\begin{equation}\label{eq21}
\lambda_0=\frac{1}{\pi\Omega}\left(\epsilon\Gamma(\frac{l}{2}+1)(\frac{l}{2}+1)\right)^{\frac{1}{\frac{l}{2}+1}},
\end{equation}  
where \eqref{eq21} is formulated by solving $\lambda_0:q^{l}(\lambda_0)=\epsilon$ with a single term in \eqref{eq20}. To refine this approximation, we have 
\begin{equation}
\lambda_{1} = \lambda_{0}-\frac{q^{l}(\lambda_0)}{q^{l'}(\lambda_0)},
\end{equation}
where $q^{l'}(\lambda_0)$ is the first order derivative of $q^{l}(\lambda_0)$. Consequently, this should give
\begin{eqnarray}\label{eq23}
\lambda_{n+1}&=& \lambda_{n}+\lambda_{n}e^{\lambda_{n}\pi\Omega}\left(\lambda_{n}\pi\Omega\right)^{-\frac{l}{2}-1}\nonumber\\
&&\cdot\left(\Gamma(\frac{l}{2}+1,\lambda_{n}\pi\Omega)+(\epsilon-1)\Gamma(\frac{l}{2}+1)\right),
\end{eqnarray}
where $\Gamma(a,b)$ is the upper incomplete gamma function. Then, following the TC formula in \eqref{eq14}, we have
\begin{equation}\label{eq24}
c^{u}(\epsilon) = \lambda_{n+1}(1-\epsilon)R.
\end{equation}
It is worth noting that several iterations in general will lead to a sufficiently accurate value of $\lambda$.


\subsection{Distribution Analysis of Random Variables}
In order to obtain the explicit expressions for OP-LB and TC-UB, we need to find the expected values of the random variables involved in \eqref{eq18} and \eqref{eq24}. Start from the distribution of $|h^{'}_{i}|^{2}$. Since the SI channel estimation error $\mathbf{\Delta}_{i}$ follows i.i.d. complex Gaussian distribution with zero mean and $\sigma^{2}_{i,\mathrm{SI}}$ variance, and the multiplied pre-processing/post-processing beamforming vectors are independent of $\mathbf{\Delta}_{i}$, following the proof in \cite{Vaze2012}, we have $h^{'}_{i}\sim\mathcal{CN}(0,\sigma^{2}_{i,\mathrm{SI}})$ and $|h^{'}_{i}|^{2}\sim\Gamma(1,\sigma^{2}_{i,\mathrm{SI}})$. Similarly, we have $|h_{i,b_k}|^2\sim\mathrm{Exp}(1)$ and $|h_{i,a_k}|^2\sim\mathrm{Exp}(1)$, so that $\psi_{i,k}\sim\Gamma(2,1)$ and $\psi^{{2}/{\alpha}}_{i,k}$ follows a generalized gamma distribution with parameters $(1,\alpha,\alpha/2)$. Thus, we have $\mathbb{E}\{|h^{'}_{i}|^{2}\}=\sigma^{2}_{i,\mathrm{SI}}$, $\mathbb{E}\{\psi_{i,k}\}=2$, and $\mathbb{E}\{\psi^{\alpha/2}_{i,k}\}=\Gamma(2+2/\alpha)/2$.

For the desired signal, the explicit distribution of the largest eigenvalue $\gamma_{i,j}$ in general is difficult to find. However, if the eigenvalue $\gamma_{i,j}$ is obtained from a Wishart matrix, based on \cite{Horn1985}, we can find its distribution from the upper and the lower bound values of $\gamma_{i,j}$. Specifically, for the condition that $N_t\leq N_r$, $\gamma_{i,j}$ is obtained directly from $\mathbf{H}_{i,j}\mathbf{H}^{H}_{i,j}$, where $\mathbf{H}_{i,j}\mathbf{H}^{H}_{i,j}$ is a Wishart matrix due to complex Gaussian distribution on $\mathbf{H}_{i,j}$; For the condition that $N_t>N_r$, we have 
\begin{eqnarray}\label{eq26}
\tilde{\mathbf{H}}_{i,j}\tilde{\mathbf{H}}^{H}_{i,j}&=&\mathbf{H}_{i,j}\tilde{\mathbf{V}}_{j}\tilde{\mathbf{V}}^{H}_{j}\mathbf{H}^{H}_{i,j}\nonumber\\
&\overset{(a)}{=}&\mathbf{H}_{i,j}\mathbf{T}_{j}\mathbf{\Lambda}_{j}\mathbf{T}^{H}_{j}\mathbf{H}^{H}_{i,j},
\end{eqnarray}
where $\tilde{\mathbf{V}}_{j}$ is formulated from the estimated SI channel $\hat{\mathbf{H}}_{j}$ similar as the way to formulate $\tilde{\mathbf{V}}_{i}$ in Sec. III. In addition, Step (a) in \eqref{eq26} is obtained by applying the eigenvalue decomposition to $\tilde{\mathbf{V}}_{j}\tilde{\mathbf{V}}^{H}_{j}$, where $\mathbf{\Lambda}_{j}=\mathrm{diag}([\underbrace{1,\ldots,1,}_{N_t-N_r}\underbrace{0,\ldots,0}_{N_r}])$, and $\mathbf{T}_{j}$ is a $N_t\times N_t$ unitary matrix whose columns are the eigenvectors of $\tilde{\mathbf{V}}_{j}\tilde{\mathbf{V}}^{H}_{j}$. Then, by applying the unitary invariance property on $\mathbf{H}_{i,j}$, we obtain that $\tilde{\mathbf{H}}_{i,j}\tilde{\mathbf{H}}^{H}_{i,j}$ is also a Wishart matrix regardless the distribution on the estimated SI channel $\hat{\mathbf{H}}_{j}$.
 
Consequently, the largest eigenvalue $\gamma_{i,j}$ of $\mathbf{H}_{i,j}\mathbf{H}^{H}_{i,j}$ is bounded by \cite{Horn1985}:
\begin{equation}\label{eq27}
\|\mathbf{H}_{i,j}\|^{2}\geq\gamma_{i,j}\geq\frac{\|\mathbf{H}_{i,j}\|^{2}}{N_t},
\end{equation}
where $\|\mathbf{H}_{i,j}\|^{2}$ follows $\Gamma(N_tN_r,1)$ distribution. Then, $\mathbb{E}\{\gamma_{i,j}\}$ can be upper bounded by $\mathbb{E}\{\|\mathbf{H}_{i,j}\|^{2}\}=N_tN_r$. Similarly, the largest eigenvalue $\gamma_{i,j}$ of $\tilde{\mathbf{H}}_{i,j}\tilde{\mathbf{H}}^{H}_{i,j}$ is bounded by \cite{Horn1985}:
\begin{equation}
\|\tilde{\mathbf{H}}_{i,j}\|^{2}\geq\gamma_{i,j}\geq\frac{\|\tilde{\mathbf{H}}_{i,j}\|^{2}}{N_r},
\end{equation}
where $\|\tilde{\mathbf{H}}_{i,j}\|^{2}$ follows $\Gamma[N_r(N_t-N_r),1]$ distribution. Then, $\mathbb{E}\{\gamma_{i,j}\}$ can be upper bounded by $\mathbb{E}\{\|\tilde{\mathbf{H}}_{i,j}\|^{2}\}=N_r(N_t-N_r)$.\footnote{For both antenna configurations, we select $\gamma_{i,j}$ upper bound to keep the same direction of inequality in \eqref{eq15}.} By inserting the obtained $\mathbb{E}\{|h^{'}_{i}|^{2}\}$, $\mathbb{E}\{\psi^{\alpha/2}_{i,k}\}$ and the upper bound of $\mathbb{E}\{\gamma_{i,j}\}$ back into \eqref{eq18} and \eqref{eq24}, we obtain the explicit OP-LB and TC-UB. 


\subsection{TC Upper Bound of HD Case}
For comparison, we provide TC-UB of HD case. Unlike TC-UB of FD case, there is no SI effect, the pre-processing beamforming vector per transmitter will be the first right singular vector of the corresponding direct link's channel matrix, e.g, as in \eqref{eq10}. Meanwhile, the post-processing beamforming vector per receiver will be formulated following the way as in \eqref{eq11}. 

To obtain the TC-UB of HD case, we first derive its OP-LB by defining the geometrical location region of its dominating interferer as
\begin{equation}\label{eq28}
R^{\mathrm{HD}}_{d}=\left[\frac{\beta_{\mathrm{HD}}|h_{i,b_{l+1}}|^2}{L^{-\alpha}\gamma_{i,j}-\frac{\beta_\mathrm{HD}|v_{i}|^2}{P}}\right]^{\frac{1}{\alpha}},
\end{equation}  
where \eqref{eq28} is formulated similar as \eqref{eq16} but has removed effects from node $a_{l+1}$ and SI. In addition, due to the half capacity (or the double target rate), $\beta_{\mathrm{HD}}$ denotes the SINR threshold for HD case and is equal to $2^{2R}-1$. Then, the corresponding OP-LB can be obtained by replacing $R_d$ in \eqref{eq17} with $R^{\mathrm{HD}}_{d}$ in \eqref{eq28}, and its approximation is given by
\begin{equation}
q^{l,\mathrm{HD}}(\lambda^{\mathrm{HD}})\approx \frac{\gamma(l,\lambda^{\mathrm{HD}}\pi\Omega^{\mathrm{HD}})}{\Gamma(l+1)},
\end{equation}
where $\lambda^{\mathrm{HD}}$ is the network intensity of HD case, and
\begin{equation}
\Omega^{\mathrm{HD}}\triangleq\beta^{\frac{2}{\alpha}}\mathbb{E}\{|h_{i,b_{k}}|^{\frac{4}{\alpha}}\}(\frac{\mathbb{E}\{\gamma_{i,j}\}}{L^{\alpha}}\})^{-\frac{2}{\alpha}}.
\end{equation}
Then, the network intensity $\lambda^{\mathrm{HD}}$ can be formulated by following \eqref{eq20} to \eqref{eq23}. Finally, the TC-UB of HD case can be obtained as
\begin{equation}
c^{u,\mathrm{HD}}(\epsilon) = 2\lambda^{\mathrm{HD}}_{n+1}(1-\epsilon)R.
\end{equation}
Here, to remove the expectations in  HD TC-UB formula, we calculate $\mathbb{E}\{|h_{i,b_k}|^{\frac{4}{\alpha}}\}=\Gamma(1+2/\alpha)$ and $\mathbb{E}\{\gamma_{i,j}\}$ follows \eqref{eq27}.



\section{Numerical Results}
In this section, we provide several experiments to examine the accuracy of our derived TC-UB, compare the proposed beamforming design with some existing methods, e.g., SVD based method as in \cite{Hunter2008}; SVD plus partial ZF based method as in \cite{Vaze2012}, and purely partial ZF based method as in \cite{Huang2012}. Here, all these three existing methods utilize one of their receiver DoFs to cancel the SI effect. In addition, we also compare our derived TC-UB for FD case with the TC-UB for HD case. We assume that the simulated FD MIMO ad-hoc network lies on a 2-D disk with a number of transceiver pairs, where their locations follow Poisson random variable with its mean equal to 200.  

\textit{Experiment 1:} In this experiment, we test the derived OP-LB and TC-UB by comparing with the corresponding simulation results. 
\begin{figure}[t] 
\begin{center}
\epsfig{figure=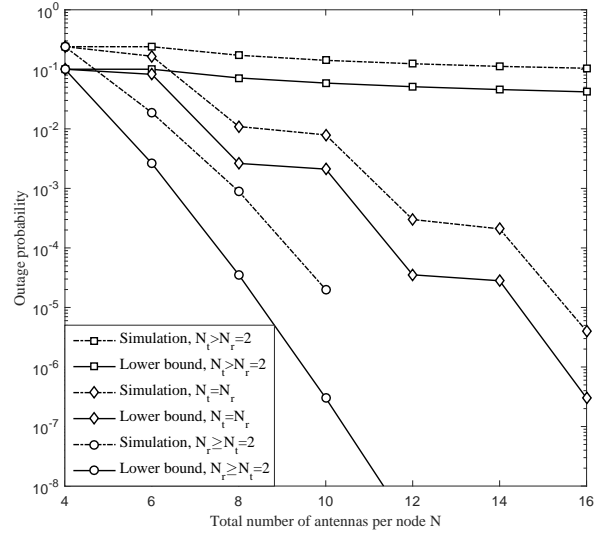,scale=0.53,angle=0}
\end{center}
\caption{Outage probability versus the total number of antennas per node, where $L=1$, $P=1$, $\alpha=4$, $\lambda=0.1$, $\sigma^{2}_{i,\mathrm{SI}}=0.1$, and $\beta=1$.}\label{F2}
\end{figure}
\begin{figure}[t] 
\begin{center}
\epsfig{figure=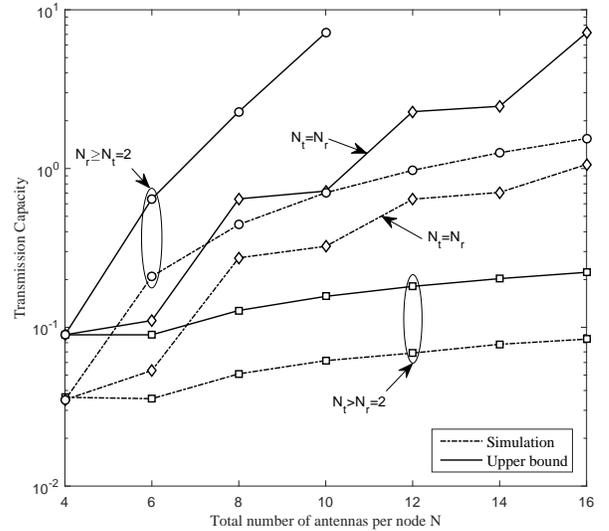,scale=0.53,angle=0}
\end{center}
\caption{Transmission Capacity versus the total number of antennas per node, where $L=1$, $P=1$, $\alpha=4$, $\epsilon=0.1$, $\sigma^{2}_{i,\mathrm{SI}}=0.1$, and $\beta=1$.}\label{F3}
\end{figure}
As shown in Fig.~\eqref{F2}, our derived OP-LB is quite close to the simulation result for different TRR configurations. However, accompanied with number of receive-antenna increasing, the performance gap is increased as well. This is due to the increased number of cancelled inter-node interferers. The more inter-node interferers being cancelled, the smaller probability that at least one interferer pair is located within the dominating interferer region. In addition, the simulation curve shows the performance with all accumulated non-dominating interferers in the network. Similar performance trend can be observed for TC-UB in Fig.~\eqref{F3}. The kinked curves for the case that $N_t=N_r$ are purely due to the specified transceiver antenna configuration.

\begin{figure}[t] 
\begin{center}
\epsfig{figure=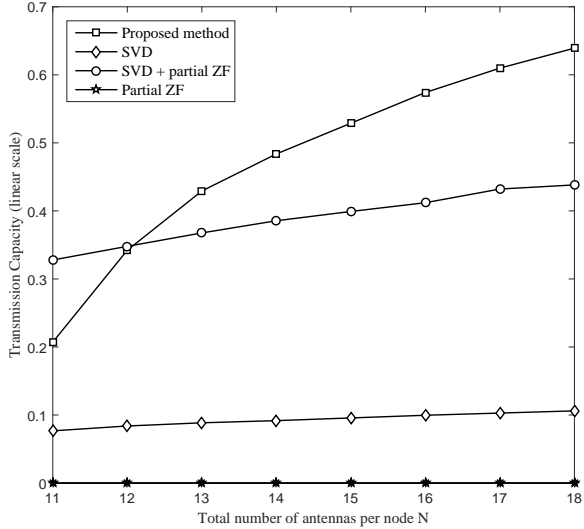,scale=0.53,angle=0}
\end{center}
\caption{Transmission Capacity versus the total number of antennas per node in comparison among different beamforming methods, where $L=1$, $P=1$, $\alpha=4$, $\epsilon=0.1$, $\sigma^{2}_{i,\mathrm{SI}}=0.1$, $\beta=1$, and $N_r=5$.}\label{F4}
\end{figure}

\textit{Experiment 2:} In this experiment, we compare the proposed beamforming design method with the above mentioned three existing beamforming strategies. As shown in Fig.~\ref{F4},
the proposed beamforming design outperforms the SVD plus partial ZF method when the total number of antenna per node is greater than 12. This is because our proposed method takes the advantage of the specified TRR configuration, i.e., $N_t>N_r$. In this case, the SI effect can be cancelled at the transmitter side, and it will leave one additional DoF at receiver side for inter-node interference cancellation. However, when $N=11$, the proposed method shows a worse performance by comparing with SVD plus partial ZF based method. This is because in this case only one DoF is left at the transmitter side and has been used for SI cancellation. In contrast, the SVD plus partial ZF method is able to use that DoF at transmitter side to boost its the desired channel link, and leave the SI to be cancelled at receiver side. In addition, the proposed method outperforms another two methods, and partial ZF method gives zero TC performance because it cannot satisfy the target OP.

\begin{figure}[t] 
\begin{center}
\epsfig{figure=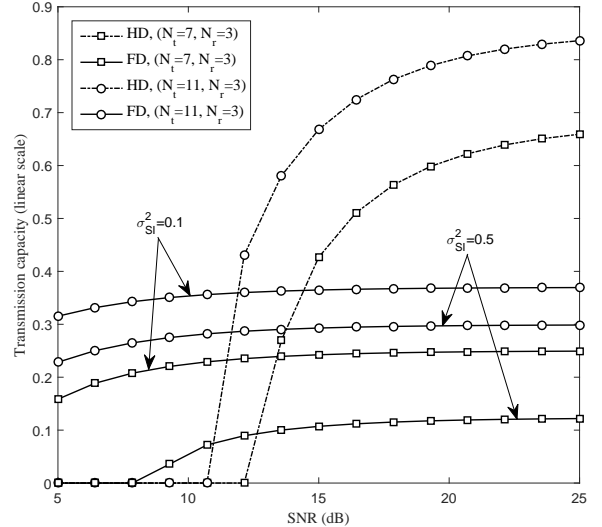,scale=0.53,angle=0}
\end{center}
\caption{Transmission Capacity versus signal to noise ratio (SNR) in comparison between FD case and HD case, where $L=1$, $\alpha=4$, $\epsilon=0.1$, and $\beta=3$.}\label{F5}
\end{figure}

\textit{Experiment 3:} In this experiment, we compare our derived TC-UB in FD mode with the TC-UB in HD mode in order to find the break-even point, where FD and HD provide the same TC performance. As shown in Fig.~\ref{F5},
with two randomly selected transceiver antenna's configurations, the FD mode outperforms the HD mode when the system works in low SNR range. For example, for the case where $N_t=7$ and $N_r=3$, the break-even point is 12.5 dB for $\sigma^2_{\mathrm{SI}}=0.5$ and 13.5 dB for $\sigma^2_{\mathrm{SI}}=0.5$. In addition, the more equipped transmit antenna, the smaller break-even point. This indicates that HD mode in this case can meet the target SINR requirement more easily, and it encounters less interference than FD mode. Moreover, as the SI channel estimation error increasing, the TC-UB in FD mode is decreasing. It is worth noting that, in low SNR range, because both HD mode and FD mode with small $N_t$ and high $\sigma^{2}_{\mathrm{SI}}$ cannot meet the $\beta=3$ requirement, their TC performances are equal to zero. 



\section{Conclusion}
In this paper, we have proposed a joint transceiver beamforming design for FD MIMO ad-hoc network to mitigate SI and partial inter-node interference effects in spatial domain. The TC-UB of the considered network has been derived in the presence of SI channel estimation error. Computer simulation and numerical results have been conducted to show that the derived TC-UB is quite close to the actual one especially when the number of receive-antenna is small, and the proposed beamforming design outperforms the existing beamforming methods when the TRR greater than one. In addition, we also compared the TC performances between the FD case and the HD case, and the result shows that TC performance in FD mode works better than that in HD mode if the system works in low SNR region.



\section*{Acknowledgement}
This work was supported by the Engineering and Physical Science Research Council (EPSRC) through the Scalable Full Duplex Dense Wireless Networks (SENSE) grant EP/P003486/1.


\bibliography{mybib}
\bibliographystyle{IEEEtran}
\end{document}